\documentstyle[11pt,aaspp4,amssym]{article}

\begin{document}

\received{}
\revised{}
\accepted{}

\lefthead{A.~V.~Sweigart \& M.~Catelan}
\righthead{The second parameter in metal-rich globulars}

\slugcomment{ApJ Letters, in press}

\title{The Second-Parameter Effect in Metal-Rich Globular Clusters}

\author{A.~V.~Sweigart and M.~Catelan}

\affil{ NASA/Goddard Space Flight Center \\
        Laboratory for Astronomy and Solar Physics, Code 681\\ 
        Greenbelt, MD 20771\\
        e-mail: sweigart@bach.gsfc.nasa.gov, catelan@stars.gsfc.nasa.gov
      }

\begin{abstract}
Recent {\it Hubble Space Telescope} observations 
have found that the horizontal branches (HB's) in the metal-rich globular
clusters NGC~6388 and NGC~6441 slope upward 
with decreasing $\bv$. Such a slope is not 
predicted by canonical HB models and cannot be produced by either 
a greater cluster age or enhanced mass loss along the red-giant branch (RGB). 
The peculiar HB morphology in these clusters may provide an important clue for 
understanding the second-parameter effect.

We have carried out extensive evolutionary calculations and numerical
simulations in order to explore three non-canonical scenarios for explaining
the sloped HB's in NGC~6388 and NGC~6441: i)~A high cluster helium 
abundance scenario, where the HB evolution is characterized by long blue loops;
ii)~A rotation scenario, where internal rotation during the RGB phase increases
the HB core mass; iii)~A helium-mixing scenario, where 
deep mixing on the RGB enhances the envelope helium abundance.
All three of these scenarios predict sloped HB's with anomalously bright RR 
Lyrae variables. We compare this prediction with the properties of the two 
known RR Lyrae variables in NGC~6388. Additional
observational tests of these scenarios are suggested.

\end{abstract}

\keywords{Stars: evolution --- stars: 
          horizontal-branch --- stars: variables: 
          other --- Galaxy: globular clusters: individual (NGC~6388, NGC~6441) 
         }

\section{Introduction}
Many years ago Sandage \& Wildey (1967) and van den Bergh (1967) discovered
that the horizontal-branch (HB) morphology in the Galactic globular clusters
(GC's) does not correlate tightly with the cluster metallicity. While theory
and observation agree that the HB morphology on average becomes redder with
increasing metallicity, there are many examples of GC's having very similar
metallicities but markedly different HB morphologies, e.g., M3 versus M13,
NGC~362 versus NGC~288. Thus some parameter besides metallicity must be
affecting the evolution of the HB stars in these clusters. Indeed,
theoretical HB models show that there are many possible ``second parameter"
candidates, e.g., the GC age, mass loss along the red-giant branch (RGB),
helium abundance $Y$, alpha-element enhancement [$\alpha$/Fe], and rotation
(Fusi Pecci \& Bellazzini 1998).

Recent {\it Hubble Space Telescope} (HST) observations have revealed that
the metal-rich GC's NGC~6388 and NGC~6441 contain a significant population
of blue HB (BHB) stars and therefore exhibit a pronounced second-parameter
effect (Piotto et al. 1997; Rich et al. 1997). From their analysis, Rich
et al. concluded that neither a greater age nor enhanced RGB mass loss due
to dynamical effects could satisfactorily account for the BHB populations
in these GC's. Understanding the origin of these BHB stars
has important implications for determining the formation history of the
Galaxy and for interpreting the
integrated spectra of the old metal-rich stellar populations in elliptical
galaxies. If, for example, age were the second parameter in these GC's,
then NGC~6388 and NGC~6441 would be among the oldest, if not the oldest,
GC's in the Galaxy.

Sweigart \& Catelan (1998) pointed out that the HB's of NGC~6388 and
NGC~6441 have a pronounced upward slope with decreasing $\bv$, with
the mean luminosity at the top of the blue tail being nearly 0.5~mag
brighter in $V$ than the well-populated red HB (RHB) clump. A similar 
upward slope is also present in the RHB itself (Piotto et al. 1997).
As we shall see, such upward sloping HB's cannot be explained within
the framework of canonical theory. More specifically, they cannot be 
attributed to differences in age or RGB mass loss---the two most
prominent second-parameter candidates. We emphasize therefore that 
these GC's may provide a crucial clue for understanding the 
second-parameter effect. 

In this {\it Letter} we will investigate the following non-canonical 
scenarios for explaining the BHB populations and, in particular, 
the sloped HB's in NGC~6388 and NGC~6441: 
\begin{enumerate}
\item {\it High $Y$ scenario}, where a high helium 
abundance at the time of GC formation leads to unusually long blue loops 
during the HB evolution; 

\item {\it Rotation scenario}, where internal rotation increases 
the HB core mass, making the HB both bluer and brighter; 

\item {\it Helium-mixing scenario}, where deep mixing along the RGB 
increases the HB envelope helium abundance, also leading to 
bluer and brighter HB stars. 
\end{enumerate}

\noindent We will explore each of these
possibilities under the simplest possible assumptions and will
show that each can produce sloped HB's resembling those 
in NGC~6388 and NGC~6441.

We begin in Sec.~2 by discussing the failure of the canonical models to 
explain the observed HB's in NGC~6388 and NGC~6441; in Sec.~3, we describe
each of the above non-canonical scenarios in more detail; 
in Sec.~4, we show that the RR Lyraes  
in NGC~6388 are brighter than field variables of
similar [Fe/H], in agreement with these scenarios. Finally, 
in Sec.~5 we suggest some observational tests to discriminate among 
these scenarios.

\section{Canonical HB Morphology}
In order to explore the HB morphology predicted by canonical models for 
NGC~6388 and NGC 6441, we have constructed a grid of HB sequences with a
main-sequence helium abundance $Y_{\rm MS} = 0.23$ and a scaled-solar 
heavy-element abundance $Z = 0.006$ (i.e., ${\rm [Fe/H]} = -0.5$), using
the stellar evolution code described by Sweigart (1997). Adopting 
the Kurucz (1992) color-temperature 
transformations and bolometric corrections, we first computed an HB 
simulation with the mean mass $\langle M_{\rm HB} \rangle$ for the
Gaussian mass distribution
and the mass dispersion $\sigma_M$ chosen to give a 
well-populated red and blue HB. This canonical simulation, 
shown in Fig.~1, predicts a completely flat HB morphology,
unlike the observed HB's in NGC~6388 and NGC~6441.
In order to explore any dependence on metallicity, we also
carried out extensive model computations for $Z = 0.002$ and
$Z = 0.01716$ ($= Z_{\sun}$), again failing to find any evidence
for sloped HB's. 

The HB simulation in Fig.~1 indicates that the BHB population in NGC~6388 
and NGC~6441 does not arise from either an older cluster age or greater 
RGB mass loss. While increasing the assumed age or RGB mass loss would move 
an RHB star blueward, it would not increase its luminosity. We conclude
therefore that canonical theory does not provide a straightforward 
explanation for the observed HB morphology in these GC's.

We have also investigated whether the slope of the RHB in NGC~6388 
and NGC~6441 could be due to differential reddening, as has been
claimed for several metal-rich GC's (e.g., 
NGC~6539:  Armandroff 1988; 
Pal~10:    Kaisler, Harris, \& MacLaughlin 1997;
NGC~6553:  Guarnieri et al. 1998).
Extensive HB simulations show that the differential reddening $\Delta E(\bv)$ 
would have to considerably exceed the 
amounts estimated by Piotto et al. (1997) in order 
to transform, say, a 47~Tuc-like RHB into the sloped HB's found in
NGC~6388 and NGC~6441. Moreover, the 
HB's in NGC~6388 and NGC~6441 are sloped in each of the four  
WFPC2 chips, which would not be expected if differential reddening 
were the cause of the sloped RHB's in the color-magnitude diagrams
(CMD's) of these GC's.

Finally, we point out that inaccuracies in the transformation 
from the WFPC2 filter system to the ``standard" Johnson-Cousins system are 
also unlikely to cause the sloped HB's in NGC~6388 and NGC~6441, since 
the instrumental CMD's  clearly show sloped HB's (Dorman 1998, priv.
comm.). If the filter transformation were responsible, one would expect 
sloped HB's in the HST CMD's for other GC's (e.g., 47~Tuc and NGC~2808) 
from Piotto et al. (1997) and Sosin et al. (1997), which is not the case.

\section{Non-Canonical HB Morphology}
The failure of canonical models to explain the observed HB morphology
in NGC~6388 and NGC~6441 suggests that some non-canonical process
may be operating in these metal-rich GC's. In this section, we will 
therefore discuss three non-canonical scenarios for producing sloped HB's, 
beginning with the possibility of a high cluster helium abundance.

\subsection{High $Y$ Scenario}
RHB stars evolve along blue loops during most of their HB lifetime. 
Normally these loops cover only a small range in 
$\bv$. For larger helium abundances, however, these blue loops can 
become considerably longer, reaching higher effective temperatures and 
deviating more in luminosity from the zero-age HB (ZAHB) (Sweigart \& Gross
1976). Thus, at least qualitatively, one would expect the HB for a sufficiently
high $Y$ to slope upward with decreasing $\bv$ (Catelan \& de Freitas 
Pacheco 1996).

To test this possibility, we have constructed a grid of 
$\approx 350$ HB sequences for 
$Y_{\rm MS} = 0.23$, 0.28, 0.33, 0.38 and 0.43 and $Z = 0.002$, 0.006 and
0.01716. For each ($Y_{\rm MS},~\,Z$) combination we 
first evolved a star up the RGB without mass loss and then through the helium 
flash to obtain a ZAHB model at the red end of the HB. Lower mass ZAHB models 
were then obtained by removing mass from the envelope of this high mass ZAHB 
model. Not surprisingly, the HB simulations for $Y_{\rm MS} = 0.23$, 0.28 and 
0.33 did not show a significant slope. However, when $Y_{\rm MS}$ was 
increased to 0.38 and 0.43, a pronounced HB slope close to that in NGC~6388
and NGC~6441 was found, as illustrated in Figs.~2a,b, respectively. 
There is also a hint of bimodality, as seen in the observed CMD's 
(Rich et al. 1997).
In these simulations we did not try to reproduce the ratio of blue to red 
HB stars. This ratio depends on the choice of $\langle M_{\rm HB} \rangle$ 
and $\sigma_M$, assumed here to be $0.02\, M_{\sun}$.  

The HB's predicted by this high $Y$ scenario are quite bright, 
implying a large pulsation period for the RR Lyrae variables and a large value 
for the number ratio $R$ of HB stars to RGB stars brighter than the mean 
RR Lyrae luminosity (Iben 1968). In addition, this scenario would imply a 
peculiar chemical enrichment history for NGC~6388 and NGC~6441, characterized 
by a very large enrichment law $\Delta Y/\Delta Z$ (cf. Shi 1995; Schramm 1997).

\subsection{Rotation Scenario}
Since the work of Mengel \& Gross (1976) it has been known that rotation during
the RGB phase can delay the helium flash. As reviewed by Renzini (1977), this
would have two consequences for the subsequent HB evolution. First, it would
increase the helium-core mass $M_{\rm c}$ and hence the HB luminosity. Second, 
it would
lead to enhanced mass loss near the tip of the RGB and thus to a smaller HB
envelope mass. The net effect would be a shift in the HB location towards
higher effective temperatures and luminosities. Thus, at least qualitatively,
one might also expect a range in rotation to produce an upward sloping HB.
Could this be the explanation for the sloped HB's in NGC~6388 and NGC~6441?

To answer this question, one needs to know how much extra mass is lost at the
tip of the RGB when $M_{\rm c}$ exceeds its canonical value. We determined
this by evolving a number of sequences up the RGB for various values of the
Reimers mass loss parameter $\eta_{\rm R}$ (cf. Reimers 1975; Fusi Pecci \&
Renzini 1976). In these sequences we turned off all helium burning, thereby 
permitting $M_{\rm c}$ to exceed its canonical value without igniting the 
helium flash. Using these sequences, we were able to determine how the final 
mass $M$ decreases with increasing $M_{\rm c}$ for each value of
$\eta_{\rm R}$. The $M - M_{\rm c}$ relations defined in this manner were 
then used to compute grids of HB sequences for $\eta_{\rm R} = 0.1$, 0.2, 
0.3 and 0.4.

Two simulations for this rotation scenario are shown in Fig.~2 for 
$\eta_{\rm R} = 0.2$ (panel c) and 0.4 (panel d). The red end of the HB
in these simulations is populated by canonical models without rotation.
The masses of these canonical models were used as the mean mass 
$\langle M_{\rm HB} \rangle$ for 
defining the mass range $M < \langle M_{\rm HB} \rangle$ covered by 
these simulations. Thus the blueward 
extension of the HB in Figs.~2c,d is driven solely by the
increase in $M_{\rm c}$ and corresponding decrease in $M$. In both Figs.~2c,d 
the HB slopes upward, especially in the $\eta_{\rm R} = 0.2$ case. 
Note, however, that the increase in $M_{\rm c}$ for the bluer HB stars
is very large ($\gtrsim 0.1\, M_{\sun}$), which, in turn, would require a very
high main-sequence rotation rate.

\subsection{Helium-Mixing Scenario}
The observed abundance variations in GC red-giant stars indicate
that these stars are able to mix nuclearly processed material from the vicinity
of the hydrogen shell out to the surface, presumably as a result of internal
rotation (Kraft 1994). In particular, the observed variations
in aluminum require the mixing to penetrate into the hydrogen shell 
(Langer \& Hoffman 1995; Cavallo, Sweigart, \& Bell 1998). 
Thus any mixing process which dredges up aluminum will also dredge up 
helium. Besides increasing the envelope helium abundance $Y_{\rm HB}$ 
such mixing would also increase the RGB tip luminosity and
thus the amount of mass loss. Consequently a red-giant star which undergoes
such helium mixing will arrive on the HB with both a higher
$Y_{\rm HB}$ and a lower mass and hence will be both hotter and 
brighter than the corresponding canonical star (Sweigart 1997, 1998). This 
suggests, at least qualitatively, that helium mixing might also lead to a 
sloped HB morphology.

To investigate this possibility, we have evolved a set of $\approx 100$ 
sequences up the RGB and through the helium flash and HB phases for
various amounts of helium mixing. Two cases were considered, namely, 
$(M,\, Y_{\rm MS},\, Z) = (0.95,\, 0.23,\, 0.006)$ and 
$(0.87,\, 0.28,\, 0.006)$, where the mass in each case corresponds to an 
age of 13~Gyr at the tip of the RGB. The $\eta_{\rm R}$ values used in 
these sequences were 0.4, 0.5 and 0.6.

Two of the helium-mixing simulations for $Y_{\rm MS} = 0.23$ are
presented in Figs.~2e,f. Each of these simulations covers the mass range 
$M < \langle M_{\rm HB} \rangle$, where $\langle M_{\rm HB} \rangle$ was
taken to be the mass of a canonical model without helium mixing. 
As the helium mixing increases, the HB tracks shift blueward, giving rise to
the pronounced upward slope evident in Figs.~2e,f. In fact, the HB slope in
Fig.~2e slightly exceeds the observed slope in NGC~6388 and NGC~6441. There
is, moreover, a hint of bimodality in this simulation. 
The helium abundance $Y_{\rm HB}$ near the instability strip in Fig.~2e
is $\approx 0.34$.

\section{Constraints from RR Lyrae Variables}
We have compared the predictions of the above simulations in the 
period-temperature diagram with the two known RRab Lyrae 
variables in NGC~6388 (Silbermann et al. 1994). 
Temperatures were determined as in Catelan, Sweigart,
\& Borissova (1998; see also Catelan 1998). The result of one such 
comparison is given in Fig.~3, where we also plot field variables 
of similar metallicity as well as V9 in 47~Tuc (Storm et al. 1994). 
The [Fe/H] values 
for the field variables are from Layden (1994); amplitudes and 
periods are taken from Blanco (1992), but in some cases from Sandage
(1990). Blazhko variables were avoided.
As one can clearly see, the variables in NGC~6388 and 
47~Tuc have considerably longer periods and thus 
substantially higher luminosities than field stars of comparable
metallicity. Moreover, the cluster variables seem to agree well 
with the displayed simulation, which in this case was taken from Fig.~2e.

\section{Discussion}
The results in this {\it Letter} demonstrate that
canonical models fail to explain the sloped HB's in NGC~6388 
and NGC~6441. This has the important implication that, at least for 
these GC's, the two most prominent second-parameter candidates---age 
and RGB mass loss---cannot account for the observed HB morphology. 
Simulations employing 
non-canonical models show that an increased initial GC helium
abundance or a dispersion in the stellar rotational velocity (leading 
to a range in $M_{\rm c}$ or $Y_{\rm HB}$) may 
account for the observed HB morphology in these two GC's. Finally, we
find that the RR Lyrae variables in NGC~6388 and 47~Tuc 
are substantially brighter than 
field variables of comparable [Fe/H]---in agreement with the 
non-canonical scenarios.

We can suggest a variety of observational tests for discriminating 
among the different non-canonical scenarios, including: 
\begin{enumerate}
\item {\it Analysis of R-ratios}: the $R$ ratios
differ widely from one non-canonical scenario to the next, being
largest for the high-$Y$ case (Figs.~2a,b). A low $R$ ratio (implying
$Y_{\rm MS} \lesssim 0.35$) for NGC~6388 and NGC~6441 would argue 
against the high-$Y$ scenario; 

\item {\it Deep photometry of NGC~6388 and NGC~6441}: this would 
determine whether the magnitude difference between the 
RR Lyraes and the turnoff, ``$\Delta V$," is larger than expected 
for a ``47~Tuc-like" age, as predicted by the non-canonical scenarios; 

\item {\it RR Lyrae survey}: 
NGC~6388 and NGC~6441 should be surveyed for 
RR Lyraes, in order to discover new variables, especially 
closer to the cluster centers, and to define better the pulsation properties 
of the known RR Lyraes. In addition, we suggest a study of the RR Lyraes in 
the metal-rich GC's NGC~6304 (Hartwick, Barlow, \& Hesser 1981; Layden 1995) 
and NGC~6652 (Ortolani, Bica, \& Barbuy 1994);

\item {\it Surface gravities of blue HB stars}: $\log\,g$
measurements could  
discriminate among the different origins of the BHB stars in
NGC~6388 and NGC~6441 (Crocker, Rood, \& O'Connell 1988) and
provide a strong constraint on whether the HB's are indeed brighter 
than predicted by the canonical models; 

\item {\it High-resolution spectroscopy of red giants}: RGB stars in 
NGC~6388 and NGC~6441 should be searched for evidence of deep mixing
(cf. Kraft 1994);

\item {\it Planetary Nebula (PN) in NGC~6441}: the PN recently 
discovered by Jacoby et al. (1997) in NGC~6441 may be an 
invaluable diagnostic tool for constraining the models. 
Helium is moderately enhanced over the solar value in 
this PN, while oxygen is more depleted than in the 
super-oxygen-poor stars which are found in some more metal-poor 
GC's and which are indicative of deep mixing (e.g., Kraft 1994). 
We speculate that this PN may be the progeny of a star that 
experienced helium mixing.

\end{enumerate}

\acknowledgments
M.C. would like to thank B. Dorman and S. Ortolani for useful
discussions. This research was supported in part by NASA grant 
NAG5-3028. This work was performed while M.C. held a 
National Research Council--NASA/GSFC Research Associateship.

\newpage

%
\begin{figure}[t]
 \plotfiddle{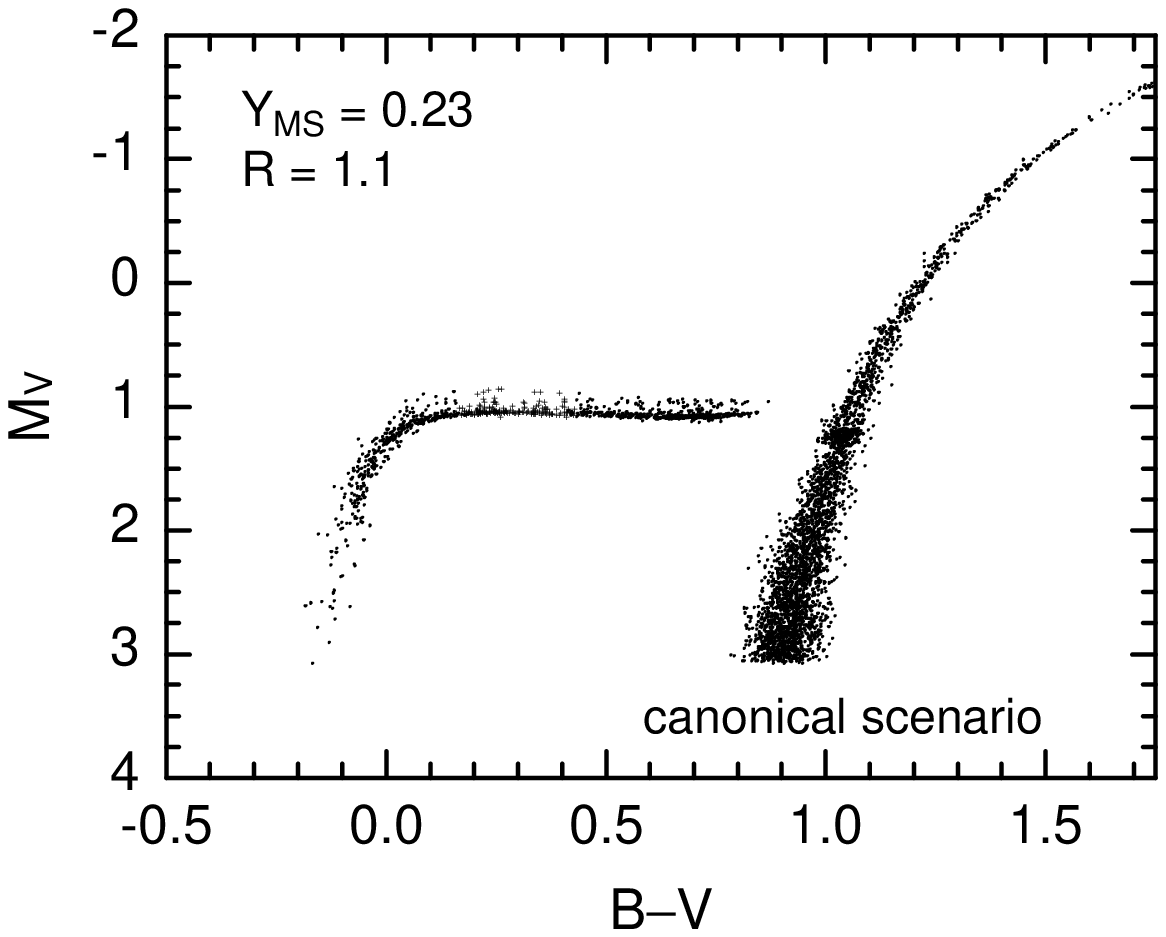}{10cm}{0}{100}{100}{-300}{-300}
 \caption[]{Canonical HB simulation for a helium abundance
 $Y_{\rm MS} = 0.23$ and a heavy-element abundance $Z = 0.006$. Note that the 
 HB is completely flat between $0.1 \lesssim \bv \lesssim 0.9$. The 
 value of the ratio $R$ 
 between HB stars and RGB stars brighter than the mean RR Lyrae luminosity 
 is indicated.}
\end{figure}

\newpage
\begin{figure}[t]
 \plotfiddle{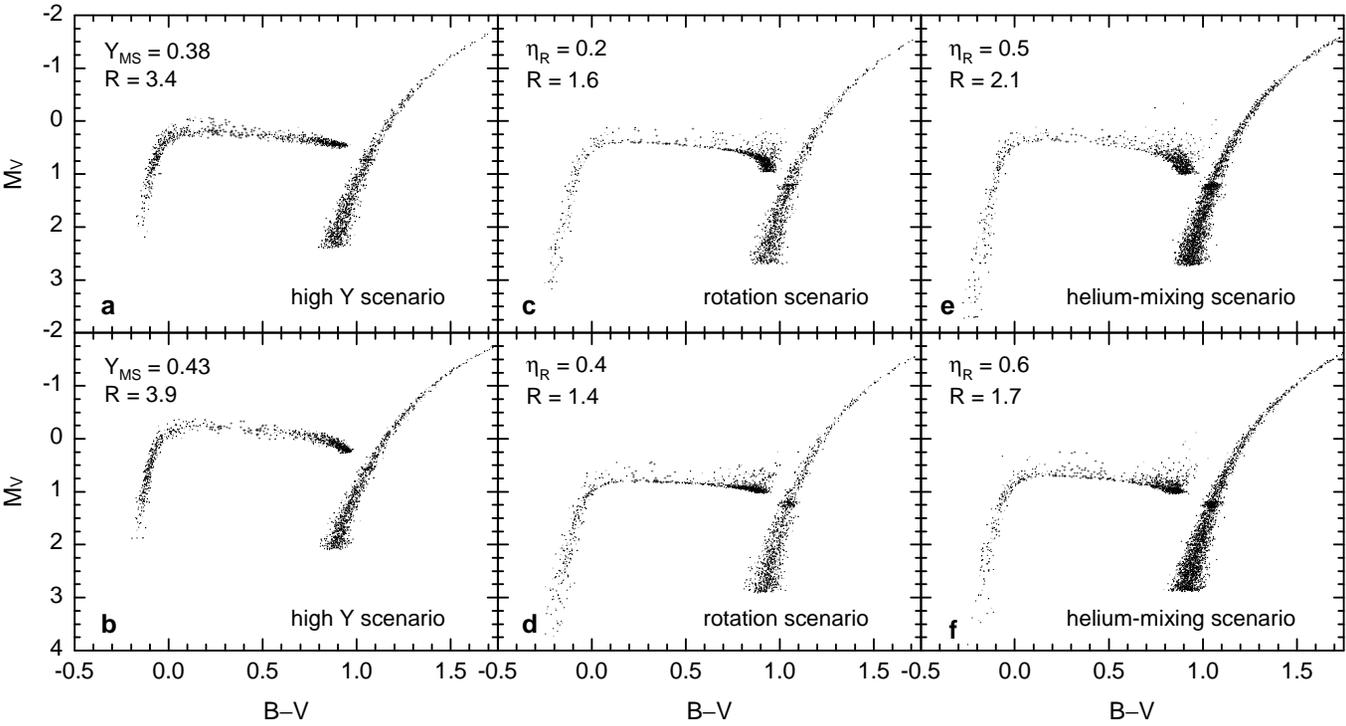}{15cm}{0}{70}{70}{-175}{-10}
  \caption[]{HB simulations for the high $Y$ scenario with 
  $Y_{\rm MS} = 0.38$ (panel a) and $Y_{\rm MS} = 0.43$ (panel b),
  the rotation scenario with the Reimers mass loss parameter 
  $\eta_{\rm R} = 0.2$ (panel c) and 0.4 (panel d), and the
  helium-mixing scenario with 
  $\eta_{\rm R} = 0.5$ (panel e) and 0.6 (panel f). The value
  of the $R$ ratio is given in each panel. Note the upward 
  slope of the HB with decreasing $\bv$.}
\end{figure}

\newpage
\begin{figure}[t]
 \plotfiddle{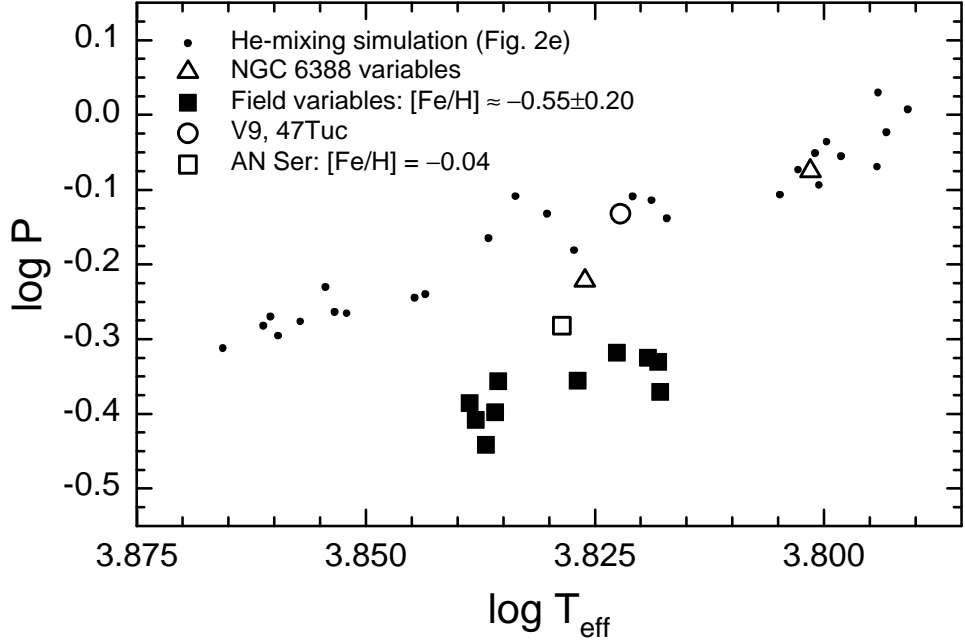}{10cm}{0}{80}{80}{-250}{-190}
 \caption[]{Period-temperature diagram for the RR Lyrae variables in NGC~6388
 ($\bigtriangleup$) compared with field variables of similar 
 metallicity ($\blacksquare$), V9 in 47 Tuc ($\bigcirc$) 
 and the very metal-rich, long-period field RR Lyrae variable AN Ser 
 ($\square$). Model predictions for the simulation in
 Fig.~2e are shown as dots.}
\end{figure}

\end{document}